\documentclass[prb,twocolumn,showpacs,amsmath,amssymb,superscriptaddress,floatfix]{revtex4}

\usepackage[english]{babel}
\usepackage{graphicx}
\usepackage{times}

\begin{document}

\title{Numerical renormalization group study of two-channel three-impurity
triangular clusters}

\author{Rok \v{Z}itko}
\affiliation{Institute for Theoretical Physics, University of G\"ottingen,
Friedrich-Hund-Platz 1, G\"ottingen, Germany}
\affiliation{J. Stefan Institute, Ljubljana, Slovenia}

\author{Janez \surname{Bon\v ca}}
\affiliation{Faculty of Mathematics and Physics, University of
Ljubljana, Ljubljana, Slovenia}
\affiliation{J. Stefan Institute, Ljubljana, Slovenia}

\date{\today}

\begin{abstract}
We study triangular clusters of three spin-$1/2$ Kondo or Anderson
impurities that are coupled to two conduction leads. In the case of Kondo
impurities, the model takes the form of an antiferromagnetic Heisenberg ring
with Kondo-like exchange coupling to continuum electrons. We show that this
model exhibits many types of the behavior found in various simpler one and
two-impurity models, thereby enabling the study of crossovers between a
number of Fermi-liquid (FL) and non-Fermi-liquid (NFL) fixed points. In
particular, we explore a direct crossover between the two-impurity
Kondo-model NFL fixed point and the two-channel Kondo-model NFL fixed point.
We show that the concept of the two-stage Kondo effect applies even in the
case when the first-stage Kondo state is of NFL type. In the case of
Anderson impurities, we consider the transport properties of three coupled
quantum dots. This class of models includes as limiting cases the familiar
serial double quantum dot and triple quantum dot nanostructures. By
extracting the quasiparticle scattering phase shifts, we compute the
low-temperature conductance as a function of the inter-impurity
tunneling-coupling. We point out that due to the existence of exponentially
low temperature scales, there is a parameter range where the stable
``zero-temperature'' fixed point is essentially never reached (not even in
numerical renormalization group calculations). The ``zero-temperature''
conductance is then of no interest and it may only be meaningful to compute
the conductance at finite temperature. This illustrates the perils of
studying the conductance in the ground state and considering thermal
fluctuations only as a small correction.
\end{abstract}


\maketitle

\newcommand{\vc}[1]{\boldsymbol{#1}}
\newcommand{\ket}[1]{|#1\rangle}
\newcommand{\deltae}{\delta_\mathrm{q.p.}^\mathrm{even}}
\newcommand{\deltao}{\delta_\mathrm{q.p.}^\mathrm{odd}}

\section{Introduction}

Quantum impurity models describe localized single impurities or impurity
clusters in interaction with conduction bands of itinerant electrons. They
appear in several different contexts in condensed matter physics: as models
for dilute magnetic impurities in metals \cite{kondo1964, gruner1974,
hewson}, as models of semiconductor quantum dots and other nanostructures
embedded between conduction leads \cite{glazman1988, kouwenhoven2001} and as
effective models within the dynamical mean-field theory of bulk correlated
electron systems \cite{georges1996}. Quantum impurity models are often
studied also for their own sake due to the fascinating and rich behavior
that they exhibit. Advances in the computational resources and improved
implementations of the numerical renormalization group (NRG) technique
\cite{wilson1975, krishna1975, yoshida1990, oliveira1994, chen1995,
krishna1980a, bulla1998, hofstetter2000, anders2005, sidecoupled,
peters2006, weichselbaum2007, bulla2008, toth2008} make possible accurate
and detailed studies of increasingly complex quantum impurity models
featuring several impurities and several conduction channels \cite{paul1996,
vzporedne, oguri2005, oguri2005phase, nisikawa2006, vzporedne2, ferrero2007,
wang2007, spincharge}.

NRG permits to calculate finite-size excitation spectra, thermodynamic
properties (such as impurity contributions to the magnetic susceptibility
and entropy), various correlation functions (in particular zero-frequency
correlators) and conductance through nanostructures both at zero temperature
and at finite temperatures \cite{bulla2008}. Such detailed knowledge about
the behavior of the system under study at different temperature scales can
be used to establish phase diagrams which delineate the parameter ranges with
characteristic properties. In more technical terms, NRG allows to determine
the possible fixed points of the model, their stability with respect to
various perturbation and crossovers between the fixed points. Depending on
the nature of the excitations, the fixed points may be classified as either
Fermi-liquid (FL) or non-Fermi-liquid (NFL) fixed points \cite{nozieres1980,
varma2002, affleck2005}. The excitation spectra of FL fixed points can be
mapped one-to-one to the spectra of free non-interacting fermions
(electrons); they have a characteristic appearance of equally-spaced lowest
lying excited states. The excitation spectra of NFL fixed points cannot be
related to non-interacting fermionic systems, they are typically more
complex and the lowest lying excited states are not equally spaced (in some
cases, however, the spectra may be described in terms of real Majorana
fermions with twisted boundary conditions; the excitation spectra may then
be given in terms of fractions \cite{maldacena1997, ye1996, ye1997,
ye1997line, sengupta1994, ye1997sf}).

Non-Fermi-liquid properties of strongly correlated materials and proposed
corresponding theoretical models have attracted the interest of the
condensed matter community due to the very uncommon situation where the
behavior of the system is radically different from what might be expected
from the nature of its elementary constituents.  The simplest models where
NFL behavior emerges are quantum impurity models such as the two-channel
single spin-$1/2$ impurity Kondo model \cite{nozieres1980, cox1996,
cox1998}. In this model, the conduction band electrons attempt to screen the
impurity moment as in the conventional Kondo effect, but as there are two
conduction channels they tend to overscreen a single spin-$1/2$ impurity.
Since a strong-coupling state with overscreening is not stable the system
ends up, instead, in a non-trivial intermediate-coupling NFL state
\cite{nozieres1980}. A mesoscopic system which exhibits the two-channel
Kondo effect has recently been experimentally demonstrated \cite{potok2007}.

In this work, we apply the numerical renormalization group techniques to
study a complex impurity model which consists of three spin-$1/2$
Kondo-model-like impurities coupled by exchange interaction so as to form a
Heisenberg ring. Two of the impurities are furthermore coupled to two
conduction bands. This system is particularly interesting in that it has a
rich phase diagram which includes fixed points known from simpler quantum
impurity models. In contrast to simpler models, however, the three-impurity
model allows to study crossovers between various non-trivial fixed points.
Spurred on by a number of experimental achievements \cite{stafford1994,
chen1994, waugh1995, jamneala2001, vidan2004, gaudreau2006,
korkusinski2007prb, schroer2007}, the interest in the transport properties
of three-impurity models recently intensified \cite{kuzmenko2003,
busser2004a, zarand2003, kuzmenko2006, oguri2006cm, torio2006, jiang2005,
ladron2006, lobos2006, jiang2007, kikoin2007, kudasov2002, savkin2005,
lazarovits2005, ingersent2005, aligia2006, tanamoto2007}. We thus also study
the conductance through a system of three coupled quantum dots described by
a related model featuring three Anderson-model-like impurity electron levels
inter-connected by tunneling coupling and connected to conduction bands by
hybridization.

In Section~\ref{secmodel} we present the model and briefly discuss
properties of known limiting cases. In Section~\ref{secregimes} we discuss
the fixed points that are expected to characterize the various parameter
ranges. Numerical results for the Kondo-like model are presented in
Section~\ref{secresults}. In this section we demonstrate the possibility of
a direct cross-over between the two-impurity Kondo model fixed point to the
two-channel Kondo model fixed point. At the same time, this result also
establishes the validity of the two-stage Kondo screening concept in the
case where the first stage of screening results in a non-Fermi-liquid fixed
point. In Section~\ref{sectqd} we discuss the transport through a triangular
triple quantum dot systems connected to two conduction leads. We emphasize
that the notion of the ``zero-temperature'' conductance is of limited
utility in systems with exponentially low energy scales, since experiments
are performed at finite temperatures.

\section{Model}
\label{secmodel}

\begin{figure}[htbp!]
\includegraphics[width=5cm,clip]{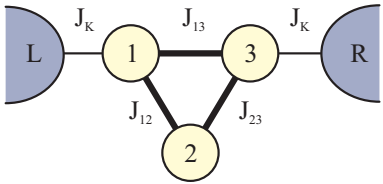}
\caption{
(Color online) Two-channel three-impurity triangular cluster.
}
\label{sistem}
\end{figure}

The three-impurity Kondo model with two conduction channels, represented
schematically in Fig.~\ref{sistem}, is described by the Hamiltonian $H =
H_\mathrm{b} + H_\mathrm{imp} + H_\mathrm{c}$, where
\begin{equation}
\begin{split}
H_\mathrm{b} &= \sum_{\nu \in \{L,R\},k,\sigma \in
\{\uparrow,\downarrow\} } \epsilon_k c^\dag_{\nu k \sigma}
c_{\nu k \sigma}, \\
H_\mathrm{imp} &= \sum_{i<j \in \{1,2,3\} } J_{ij} \vc{S}_i \cdot \vc{S}_j,\\
H_\mathrm{c} &= J_K \left(  \vc{S}_1 \cdot \vc{s}_L + \vc{S}_3 \cdot
\vc{s}_R \right).
\label{eq1}
\end{split}
\end{equation}
The two ($\nu=L$ and $\nu=R$) conduction bands are assumed to have linear
dispersion, $\epsilon_k=Dk$, where the dimensionless wave-number $k$ ranges
from $-1$ to $1$ so that the bandwidth is $2D$. The impurities are described
by the spin-$1/2$ operators $\vc{S}_i$ with $i=1,2,3$ and $\vc{s}_\nu$ is
the conduction-band spin density at the position of impurity 1 (3) for
$\nu=L$ ($R$), 
\begin{equation}
\vc{s}_\nu = \sum_{\alpha\alpha'} 
\left( \frac{1}{\sqrt{N}} \sum_k c^\dag_{\nu k\alpha} \right)
\left( \frac{1}{2} \boldsymbol{\sigma}_{\alpha\alpha'} \right)
\left( \frac{1}{\sqrt{N}} \sum_{k'} c_{\nu k'\alpha'} \right),
\end{equation}
where $N$ is the number of conduction-band states, $c^\dag_{\nu k \alpha}$
and $c_{\nu k \alpha}$ are creation and annihilation operators for electron
in band $\nu$ with spin $\alpha \in \{ \uparrow, \downarrow \}$, and
$\boldsymbol{\sigma}_{\alpha\alpha'}$ is the vector of Pauli matrices
$(\sigma^x_{\alpha\alpha'}, \sigma^y_{\alpha\alpha'},
\sigma^z_{\alpha\alpha'})$. $J_K$ is the antiferromagnetic Kondo exchange
constant. Finally, $J_{ij}$ parameterize the inter-impurity exchange
interaction. We reduce the parameter space by considering only models with
left-right mirror symmetry (parity), i.e. $J_{12} = J_{23}$. While the
parity-breaking is a relevant perturbation at some of the fixed points
\cite{affleck1995}, its effect will not be studied in much detail in this
work. We parameterize the exchange constants by
\begin{align}
J_{13} &= J_0 \sin \left( \beta \pi/2 \right), \\
J_{12} = J_{23} &= J_0 \cos\left( \beta \pi/2 \right).
\end{align}
Parameter $\beta$ thus parameterizes the asymmetry (ratio) between the
exchange coupling between the impurities 1 and 3 in the upper arm of the
triangle and the exchange coupling between the side-coupled impurity 2 and
the impurity 1 (or 3). The special cases are:
\begin{itemize}
\item $\beta=1$, which corresponds to the two-impurity Kondo model 
(plus one totally decoupled impurity);

\item $\beta=0$, which corresponds to a linear chain of three impurities; 

\item $\beta=1/2$ which corresponds to a symmetric antiferromagnetic Heisenberg
ring, which is a magnetically frustrated system with two degenerate doublets
in the ground state.
\end{itemize}
Parameter $J_0$ sets the overall scale of the inter-impurity exchange
coupling.

In the two-impurity Kondo model \cite{jayaprakash1981} (i.e. $\beta=1$
limit), there is a double Kondo screening regime for low inter-impurity
exchange interaction $J=J_0$ and an inter-impurity-singlet regime for high
$J$ \cite{jones1987, jones1988, jones1989, jones1989prb, affleck1992,
affleck1995, sakai1992, campo2004, zarand2006, zhu2006}. These phases are
separated by a quantum phase transition \cite{vojta2003qpt, vojta2006} at
$J=J^*_\mathrm{2IK} = c T_K^{(1)}$ where $T_K^{(1)}$ is the Kondo
temperature for a system of a single impurity coupled to a single conduction
channel with the same $J_K$ (the proportionality constant $c$ is of order
$1$; often quoted value is $\sim 2.2$ obtained in the early NRG studies
\cite{jones1988}, however this particular value is not universal and true
$c$ depends on the details of the model and on the chosen definition of the
Kondo temperature).  Exactly at the transition point, the system has a
non-Fermi-liquid ground state and exhibits quantum criticality. This state
is, however, unstable and for $J-J^*_\mathrm{2IK} \neq 0$ the systems flows
to either of the two possible FL fixed points. We denote this NFL fixed
point by 2IK. Recently, the two-impurity models were reexamined and it was
shown that this fixed point is robust with respect to parity, particle-hole
symmetry breaking and various other asymmetries \cite{zarand2006}.

The system of three impurities in series (i.e. $\beta=0$ limit) has been
studied in Ref.~\onlinecite{flnfl3}. This system has a non-Fermi-liquid
ground state of the same type as the two-channel Kondo model (2CK)
\cite{cox1998, affleck2ck1992, affleck1995, kuzmenko2003, pang1991,
pang1994, ludwig1991, sengupta1996, maldacena1997, vondelft1998, zarand2002,
affleck1991over, coleman1995prl, coleman1995, andrei1995}. For low
inter-impurity exchange interaction $J$, the local moment screening occurs
in two stages: at the higher Kondo temperature $T_K^{(1)}$ the local moments
on impurities 1 and 3 are screened, while the local moment on impurity 2 is
screened at an exponentially reduced second Kondo temperature $T_K^{(2)}$
\cite{flnfl3, tripike}. For high $J$, the three spins first lock into an
antiferromagnetic spin-chain at $T \sim J$ and the collective spin-$1/2$
undergoes Kondo screening at some lower temperature $T_\mathrm{2CK}$ which
depends non-monotonically on $J$ \cite{flnfl3, kolf2007}. The
low-temperature two-channel Kondo (2CK) fixed point is stable with respect
to particle-hole symmetry breaking \cite{kusunose1996}, but it is unstable
with respect to parity breaking \cite{cox1998, affleck1995,
pustilnik2ck2004, flnfl3}.

For $\beta=0.5$ and sufficiently large $J_0$, the three impurities behave as
a frustrated antiferromagnet with two degenerate ground state doublets at
temperatures on the scale of $J_0$. The symmetry is broken by the coupling
to the leads as there are only two conduction channels. It should be noted
that in the more symmetric case of three conduction channels frustration
induces a new type of non-Fermi-liquid behavior \cite{paul1996,
ferrero2007}.

\section{Expected regimes}
\label{secregimes}

\begin{figure}[htbp!]
\includegraphics[width=8cm,clip]{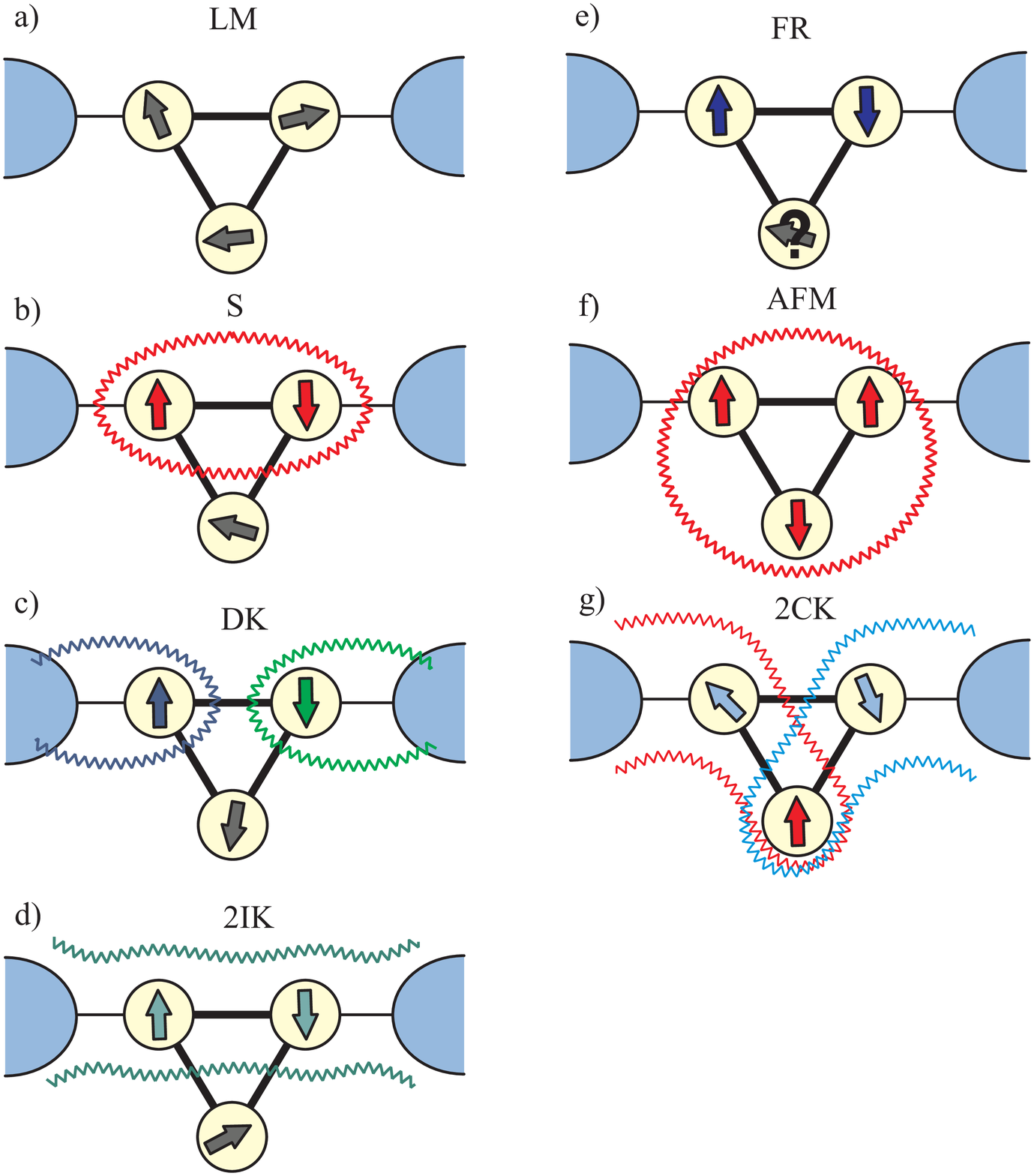}
\caption{
(Color online) Schematic representations of the possible fixed points.
}
\label{fps}
\end{figure}

The behavior of the system at various temperatures and inter-impurity
coupling strengths is governed by the proximity to one of the following
fixed points (see schematic representations in Fig.~\ref{fps}):
\begin{enumerate}
\item[a)] three independent local moments, LM,

\item[b)] inter-impurity singlet (plus a decoupled spin-$1/2$ local moment), S,

\item[c)] Kondo screening with $\pi/2$ phase shifts (plus a decoupled
spin-$1/2$ local moment), DK,

\item[d)] two-impurity Kondo model non-Fermi-liquid fixed point (plus a
decoupled spin-$1/2$ local moment), 2IK,

\item[e)] frustrated antiferromagnetic Heisenberg ring, FR,

\item[f)] antiferromagnetic spin chain with $S=1/2$, AFM (two different fixed
points exist, depending on $J_{12} > J_{13}$ or $J_{12} < J_{13}$),

\item[g)] two-channel  spin-$1/2$ Kondo model non-Fermi-liquid fixed point,
2CK.
\end{enumerate}
It should be noted that we have restrained ourselves to the fixed points
which occur in the mirror (parity) and particle-hole symmetric case. In
generic model, some of these fixed points are extended into lines or planes
of fixed points. In particular, there appears a plane of Fermi-liquid fixed
points (plus a decoupled spin-$1/2$ local moment) parameterized by two
continuous quantities (phase shifts in even and odd scattering channel), of
which the fixed points S and DK are special cases. There is furthermore a
new plane of Fermi-liquid fixed points that is also parameterized by the two
phase shifts but this time there is no decoupled local moment. The first
plane (P1) occurs for $\beta=1$ and the second one (P2) for $\beta \neq 1$;
for $\beta \lesssim 1$, the crossover from P1 to P2 occurs by a Kondo-like
screening of the nearly decoupled local moment.

\begin{figure}[htbp!]
\includegraphics[width=8cm,clip]{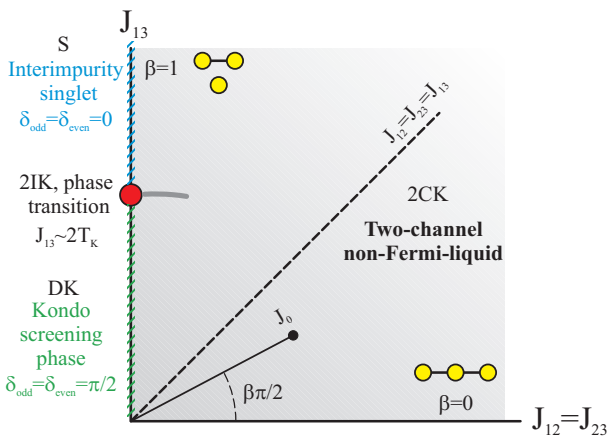}
\caption{
(Color online) Schematic phase diagram of the triangular three impurity
Kondo model at low temperatures.
}
\label{skica}
\end{figure}

In Fig.~\ref{skica} we present a schematic ``phase diagram'' of the system
in the $(J_{12}, J_{13})$ [or, equivalently, $(J_0,\beta)$] plane at very
low ($T \to 0$) temperatures. Along the $\beta = 1$ line we find the S and
DK Fermi-liquid fixed points, separated by the 2IK quantum phase transition.
Along the $\beta=0$ line as well as in the rest of the plane, the system
always ends up in the 2CK fixed point. This is a special property of the
parity-symmetric case (in the fully generic case, only P1 and P2 fixed
points are stable). For $T>\max(J_{12}, J_{13})$, the system is in the local
moment LM regime, while for $T \lesssim \min(J_{12}, J_{13})$ the system is
described by FR or one of the two AFM fixed points, depending on the values
of $J_{12}$ and $J_{13}$. In the region of the parameter plane in close
vicinity of the 2IK fixed point, i.e. for $J_0 \sim J^*_\mathrm{2IK}$ and
$\beta \sim 1$, the system first approaches the 2IK fixed point, then
crosses over into the stable 2CK fixed point.

We emphasize that the 2CK fixed point dominates the low-temperature phase
diagram, i.e. it is the stable fixed point for all $\beta \neq 1$.  The
values of parameters $J_0$ and $\beta$ affect only the way in which this
fixed point is approached. A similar reasoning as in the $\beta=0$ case
\cite{flnfl3} can also be applied to the general model. At some low enough
energy scale, the impurity cluster and the nearby conduction band electrons
effectively form a spin-$1/2$ object. This object is very localized in the
large-$J_0$ limit (when the three impurities align antiferromagnetically
into a spin-$1/2$ state) or extended in the small-$J_0$ limit (in which case
the composite spin-$1/2$ object is formed by the spin-$1/2$ of the impurity
2 and two collective spin-singlet states each consisting of an impurity and
electrons which screen its spin). In both cases the coupling of this
spin-$1/2$ object to the rest of the system (conduction band electrons at
still lower energies in each of the lead) is antiferromagnetic; for large
$J_0$ this is explicit, since $J_K$-terms in the Hamiltonian describe an AFM
exchange interaction, while for small $J_0$ this is expected by analogy with
the two-stage Kondo effect in side-coupled geometry \cite{vojta2002,
cornaglia2005tsk, sidecoupled}. Thus the effective model takes the form of a
single-impurity two-channel Kondo model. Since we consider
reflection-symmetric models in this work, the exchange coupling to both
channels are the same, thus the effective model with flow to the stable 2CK
NFL fixed point. Of course, in the absence of reflection symmetry, the
system would eventually end up in a Fermi-liquid fixed point (in the P2
plane).

The presence of all the enumerated fixed points and the proposed features of
the phase diagram are fully supported by the results of the numerical
renormalization group calculation which are detailed in the following
sections.

\section{Numerical results}
\label{secresults}

Calculations have been performed using the ``NRG Ljubljana'' package
\cite{nrglj}. The impurity model, Eq.~\eqref{eq1}, is particle-hole
symmetric; in fact, it has a larger $\mathrm{SU}(2)$ isospin symmetry of
which the particle-hole transformation symmetry is merely a subgroup
\cite{jones1988, affleck1995, spincharge}. We performed all calculations
taking explicitly into account spin $\mathrm{SU}(2)$, isospin
$\mathrm{SU}(2)$ and mirror $\mathrm{Z}_2$ symmetry groups \cite{wilson1975,
krishna1980a, krishna1980b, silva1996, campo2005, spincharge, bulla2008}. We
have used the discretization scheme described in Ref.~\onlinecite{campo2005}
with the discretization parameter $\Lambda=4$. Averaging over four values of
the twist parameter $z$ has been used \cite{oliveira1994}. The NRG
truncation cutoff was set at the cutoff energy of $E_\mathrm{cutoff}=9
\omega_N$, where $\omega_N$ is the characteristic energy scale at the $N$-th
NRG iteration, or at most 4000 states (which corresponds to approximately
32000 states taking into account the degeneracies). To prevent systematic
errors, care is taken not to truncate within a cluster of nearby almost
degenerate states. In all calculations presented in this article, we have
used $\rho J_K=0.2$ which corresponds to the Kondo temperature $T_K^{(1)}
\approx 0.003 D$ in the single-impurity Kondo model with the same parameter
$J_K$. (Here we use Wilson's definition of the Kondo temperature, i.e. for
$S=1/2$ Kondo model one has $k_B T_K \chi(T_K)/(g\mu_B)^2=0.701$ or $k_B T_K
\chi(0)/(g\mu_B)^2=0.103$. This definition is commonly, albeit not
exclusively, used in NRG literature.)

\begin{figure*}[tb]
\includegraphics[width=18cm,clip]{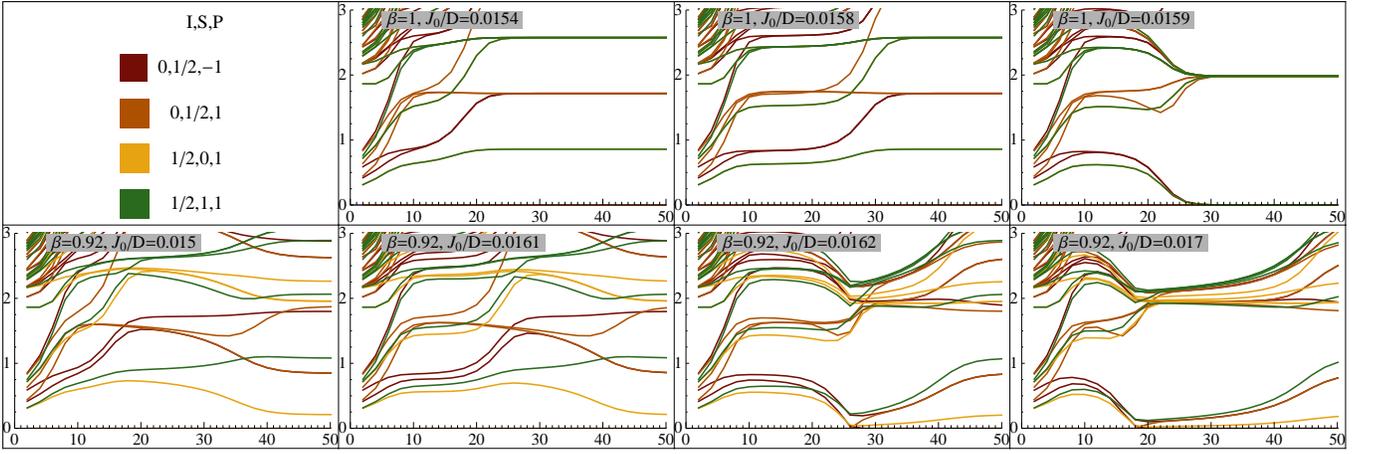}
\caption{(Color online) NRG eigenvalue renormalization flow diagrams for
even iteration numbers $N$. The states are indexed by the total isospin,
total spin and parity quantum numbers.
}
\label{fig2_c3}
\end{figure*}

We first ascertain the presence of the expected fixed points by calculating
the finite-size excitation spectra. These spectra can be represented in the
form of the ``renormalization flow diagrams'', some of which are shown in
Fig.~\ref{fig2_c3}. Flow diagrams show the NRG eigenvalue spectrum in units
of the characteristic energy (or, equivalently, temperature) scale $\omega_N
\propto \Lambda^{-N/2}$ as a function of the NRG iteration number $N$ (see
Refs.~\onlinecite{wilson1975, krishna1980a} and \onlinecite{bulla2008}). We
join the points by lines for easier visualization and interpretation. The
colors (shades of gray) correspond to different sets of quantum numbers for
the total isospin ($I$), total spin ($S$) and parity ($P$). The system is
said to be near some fixed point when the eigenvalues do not change much
between successive iterations (i.e. the lines are horizontal), while
crossovers correspond to transitions between such regions.

For $\Lambda=4$, the single-particle eigenvalues which can be combined to
give the Fermi liquid fixed point eigenspectra of a single Wilson chain with
no impurities are \cite{wilson1975, krishna1980a}
\begin{equation}
\begin{split}
\eta^*_j &= 0.8589029, 3.99452, \Lambda^2, \Lambda^3, 
\ldots, \quad N\, \text{odd}; \\
{\hat \eta}^*_j &= 1.983281, 7.999996, \Lambda^{5/2}, \Lambda^{7/2},
\ldots, \quad N\, \text{even},
\end{split}
\end{equation}
for odd and even iteration number $N$, respectively. In the first row of
Fig.~\ref{fig2_c3} we show the NRG eigenvalue flow for $\beta=1$ for three
values of $J_0$. For $J_0/D=0.0154$, we have $J_0 < J^*_\mathrm{2IK}$ and we
expect a flow to the Fermi-liquid double-Kondo (DK) fixed point with
$\deltae=\deltao=\pi/2$ quasiparticle scattering phase shifts. The $\pi/2$
phase shifts in both channels imply that the excitation spectrum of
even-length Wilson chains correspond to that of odd-length non-interacting
chain. Indeed, the lowest excitation energies are $\eta^*_0, 2\eta^*_0,
3\eta^*_0, \ldots$ to high accuracy. The quantum numbers are, however,
different from those of the related two-impurity problem due to the presence
of a decoupled spin-$1/2$ impurity (which also implies an additional
two-fold degeneracy of all levels). For $J_0/D=0.0159$, we have $J_0 >
J^*_\mathrm{2IK}$, which corresponds to the flow to the Fermi-liquid
inter-impurity singlet (S) phase with $\deltae=\deltao=0$ phase shifts. The
excitation spectrum of even-length Wilson chains corresponds to that of
even-length non-interacting chain with lowest excitation energy ${\hat
\eta}^*_0$. Finally, the unstable fixed point for $N=10,{\ldots},20$ at $J_0
\sim J^*_\mathrm{2IK} \approx 0.0158 D$ is the 2IK NFL fixed point with the
energy spectrum (after suitable rescaling) described by the fractions $3/8,
1/2, 7/8, 1, \ldots$ \cite{affleck1995, ye1997line}.

The second row in Fig.~\ref{fig2_c3} shows the flow diagrams for constant
$\beta=0.92$ for a range of values of $J_0$. In all cases the system ends up
in the same stable fixed point, which is the 2CK NFL fixed point with energy
spectrum (after suitable rescaling) $1/8, 1/2, 5/8, 1, 1+1/8, \ldots$, as
predicted by the boundary conformal field theory approach to the 2CK problem
\cite{affleck2ck1992}. Note the similarities in the flow diagrams in the
first and second line, especially at high and intermediate temperatures (up
to $N \sim 20$). At lower temperatures, the coupling to the impurity 2
eventually drives the system to the 2CK fixed point for any value of $\beta
\neq 1$ and $J_0$.

\begin{figure*}[htbp!]
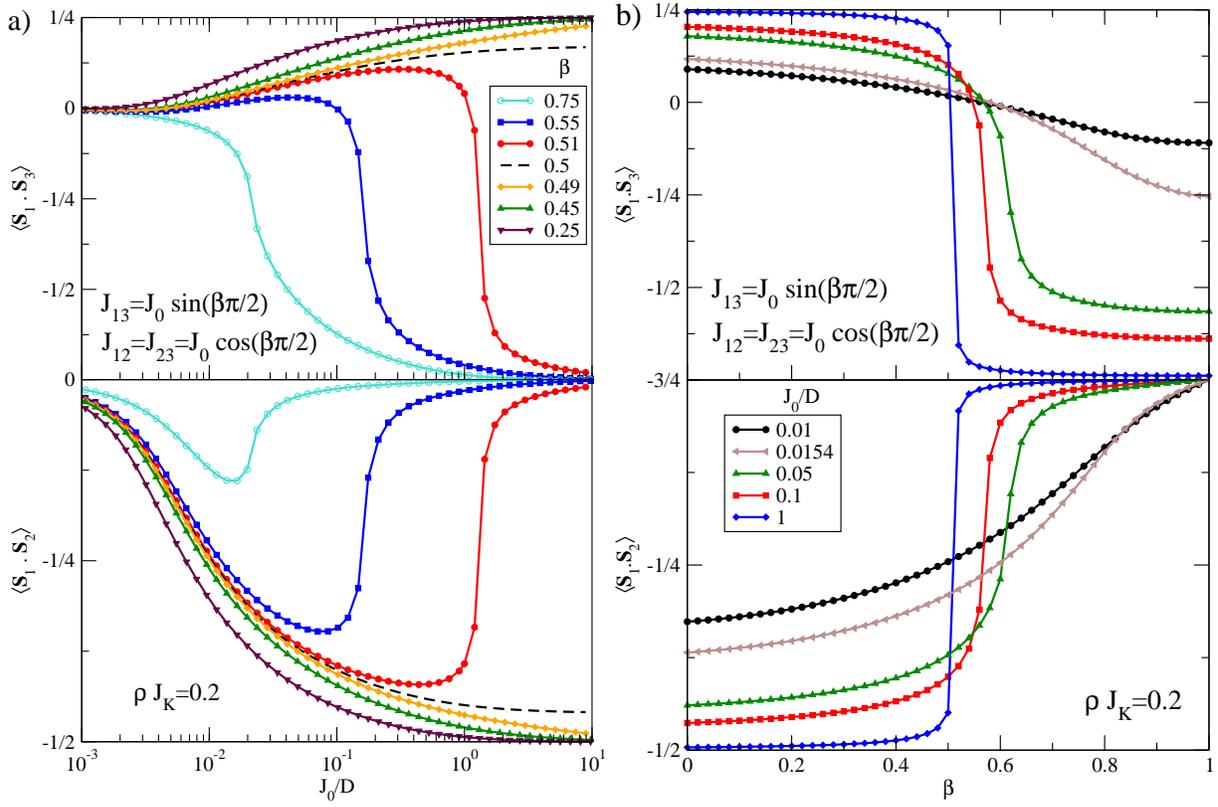

\includegraphics[width=8cm,clip]{fig5.eps}
\includegraphics[width=8cm,clip]{fig6.eps}
\caption{
(Color online) Spin-spin correlation functions between the impurities in the
upper arm of the triangle ($\vc{S}_1 \cdot \vc{S}_3$) and between the
side-coupled impurity and one of the impurities in the upper arm ($\vc{S}_1
\cdot \vc{S}_2$). Note that $\vc{S}_1 \cdot \vc{S}_2 = \vc{S}_3 \cdot
\vc{S}_2$ due to reflection symmetry. The correlations shown are those at
$T=0$; in the major part of the parameter space, the final values are
established on the scale $\sim \min(J_{13}, J_{12})$ which is typically much
higher than the Kondo temperature scales.
}
\label{fig2_b}
\end{figure*}

Basic information about the magnetic correlations within the three impurity
cluster may be obtained by considering the spin-spin correlations $c_{ij} =
\langle \vc{S}_i \cdot \vc{S}_j \rangle$ at zero temperature. These provide
insight in the competition between the inter-impurity interactions $J_{ij}$
and the impurity-lead Kondo exchange interaction $J_K$. In Fig.~\ref{fig2_b}
we plot $c_{13}$ and $c_{12}$ both as a function of the interaction strength
$J_0$ and as a function of the asymmetry ratio $\beta$. Note that if the
impurities were decoupled from the conduction channels (i.e. for $J_K/J_0
\equiv 0$), the spin correlation at zero temperature would depend only on
the parameter $\beta$, not on $J_0$. The behavior of the decoupled cluster
approximates the properties of the impurity system in the large $J_0$ limit,
see the $J_0/D=1$ plots in Fig.~\ref{fig2_b}b. At $\beta=1/2$, the decoupled
cluster has $C_{3v}$ symmetry, thus $c_{12} = c_{13}$. The two degenerate
spin-doublets at this point are
\begin{equation}
\begin{split}
\ket{ a_\sigma } &= 1/\sqrt{2} \left( | \uparrow,\sigma,\downarrow \rangle
- | \downarrow,\sigma,\uparrow \rangle \right), \\
\ket{ b_\sigma } &= 1/\sqrt{6} \left( | \downarrow,\uparrow,\sigma \rangle
+ | \uparrow,\downarrow,\sigma \rangle - 2 | \uparrow, \uparrow, \bar{\sigma} \rangle
\right).
\end{split}
\end{equation}
For an equal mixture of these two states, we expect $c_{12} = c_{13} =
-1/2$. For $\beta \neq 1/2$ the degeneracy is lifted. For $\beta > 1/2$, the
decoupled impurities are in state $\ket{a}$, which corresponds to a local
singlet state between impurities 1 and 3 ($c_{13}=-3/4$), while the impurity
2 is decoupled ($c_{12}=0$). For $\beta < 1/2$, the decoupled impurities are
in state $\ket{b}$, which corresponds to a rigid antiferromagnetic spin
chain ($c_{12}=-1/2$, $c_{13}=1/4$).

In the full model ($J_K \neq 0$), the degeneracy between the doublets is
lifted even at $\beta=1/2$ by the coupling to the channels. Curiously, in
the limit $J_0 \gg J_K$ , the correlations $c_{12}$ and $c_{13}$ are not
only different, but they even have opposite signs. This is explained by
Fig.~\ref{fig2_b}b: the point where $c_{12} = c_{13}$ is shifted from the
value of $\beta=1/2$. We also note that at large $J_0$, the transition
between $\ket{a_\sigma}$ and $\ket{b_\sigma}$ impurity ground states becomes
increasingly sharp, thus a small change in $\beta$ leads to an abrupt change
in the spin correlations.  The spin-spin correlations curves at $\beta=1/2$
play the role of separatrix between two different limiting regimes (dashed
curves in Fig.~\ref{fig2_b}a. In the other limit of $J_0 \ll J_K$, the spin
correlations tend to zero due to the magnetic screening by conduction
electrons.  It may also be noted that the ``side-coupled'' impurity 2 is
aligned antiferromagnetically with respect to the ``directly-coupled''
impurities 1 and 3 for any values of parameters $J_0$ and $\beta \neq 1$.
Since antiferromagnetic exchange is a relevant perturbation, this implies
that the local moment on impurity 2 will always be screened (except for
$\beta=1$).

At $\beta = 1$, we have studied $c_{13}=\langle \vc{S}_1 \cdot \vc{S}_3
\rangle$ at the 2IK critical point, $J_0 = J^*_\mathrm{2IK}$, which is
expected to be equal to $-1/4$ due to a degeneracy between one singlet and
one triplet state \cite{jones1987, gan1995}. Two general remarks concerning
NRG calculations are in order at this point. The first concerns the value of
the discretization parameter $\Lambda$. While even very high values of
$\Lambda$ typically lead to results which are qualitatively correct (in the
single-impurity Anderson model, one can obtain decent magnetic
susceptibility curves even at surprisingly high $\Lambda=40$), some details
depend crucially on taking the $\Lambda \to 1$ limit. The position of the
quantum phase transition as a function of $J$ and the value of $\langle
\vc{S}_1 \cdot \vc{S}_3 \rangle$ in the two-impurity Kondo model are one
such example. At $\Lambda=4$, we find $J^*_\mathrm{2IK}/D \approx 0.01585$
and $c_{13} \approx -0.255$, while at $\Lambda=2$ we find
$J^*_\mathrm{2IK}/D \approx 0.01497$ and $c_{13} \approx -0.253$. The
convergence to the expected value of $c_{13}=-1/4$ is thus relatively slow.
The second remark concerns the averaging over the twist parameter $z$ (the
``$z$-trick''). We find that the precise value of the exchange parameter $J$
at the critical point depends slightly on the value of $z$ at constant
$\Lambda$, the more so as $\Lambda$ is increased. To accurately study the
detailed properties of the model in the vicinity of critical points, the
averaging over $z$ is thus better to be avoided and $\Lambda$ should be kept
small. Finally, it may also be noted that $c$ in $J_\mathrm{2IK}^* = c
T_K^{(1)}$ is here approximately equal to $5.0$, not $2.2$ (still using
Wilson's definition of the Kondo temperature), see Section II.

\begin{figure}[htbp!]
\includegraphics[width=8cm,clip]{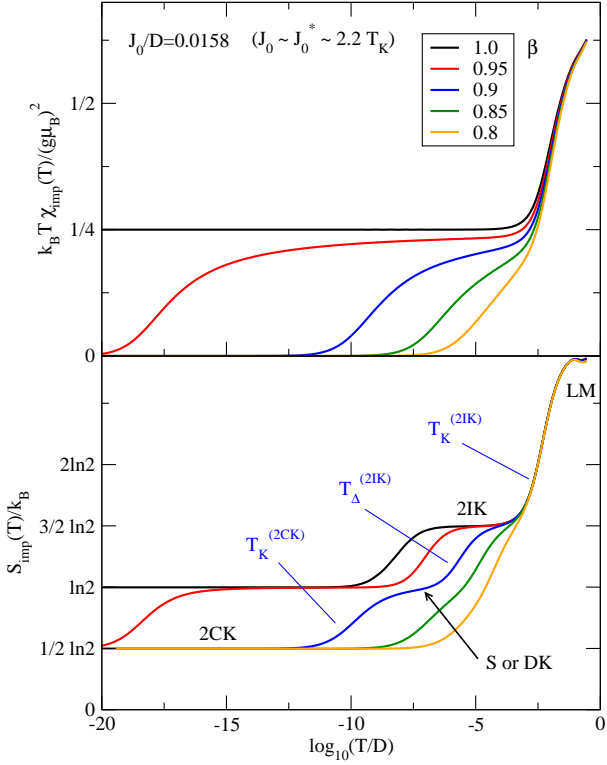}
\caption{
(Color online) Impurity cluster contribution to the magnetic susceptibility
and entropy for a range of values of $\beta$ at fixed $J_0$.
} \label{fig2_d2}
\end{figure}

We now focus on the vicinity of the 2IK fixed point, i.e. to the region
$\beta \sim 1$, $J_0 \sim J^*_\mathrm{2IK}$. For $\beta = 1$, the system
crosses over from LM to 2IK fixed point at the two-impurity Kondo
temperature $T_K^{(\mathrm{2IK})}$. Since this fixed point is not stable,
the system then crosses over at some lower temperature
$T_\Delta^\mathrm{(2IK)} \propto (J-J^*_\mathrm{2IK})^2 /
T_K^\mathrm{(2IK)}$ to either S or DK fixed point, depending on whether $J_0
> J^*_\mathrm{2IK}$ or $J_0 < J^*_\mathrm{2IK}$. This can be observed in
Fig.~\ref{fig2_d2} (black curve, $\beta=1$), where we plot two thermodynamic
quantities, the impurity contribution to the magnetic susceptibility $k_B T
\chi_\mathrm{imp} (T)/(g\mu_B^2) = \mu_\mathrm{eff}$, and the impurity
contribution to the entropy $s_\mathrm{imp}(T)/k_B=\ln \nu_\mathrm{eff}$ as
a function of the temperature. Note that the first quantity may be
interpreted as the effective impurity cluster magnetic moment
$\mu_\mathrm{eff}$, while the second one can be related to the effective
number of degrees of freedom of the cluster, $\nu_\mathrm{eff}$. At the
Kondo temperature $T_K^\mathrm{(2IK)}$, two local moments are screened,
while the third decoupled spin remains free; correspondingly,
$\mu_\mathrm{eff}$ is reduced from $3/4$ to $1/4$. In the two-impurity Kondo
effect, the NFL fixed point is associated with $1/2 \ln 2$ residual entropy,
thus the entropy goes from the LM value of $3\ln 2$ to $\ln2 +1/2 \ln2$. The
residual entropy of $1/2 \ln 2$ is released at $T_\Delta^\mathrm{(2IK)}$,
when the system crosses over to the FL stable fixed point. Since impurity 2
is completely decoupled at $\beta=1$, there is a residual spin-$1/2$ local
moment with $\ln 2$ entropy.

Keeping $J_0$ constant and decreasing $\beta$, we weakly couple the impurity
2 to the rest of the system, see Fig.~\ref{fig2_d2}. One effect is the
increased crossover temperature $T_\Delta^\mathrm{(2IK)}$. More importantly,
the local moment on the impurity 2 is now screened at some lower temperature
$T_K^\mathrm{2CK}$. This can be observed both in $\mu_\mathrm{eff}$ curves,
where the effective moment is reduced from $1/4$ to $0$, and in the
effective entropy which is reduced by $1/2 \ln 2$, which is a characteristic
value for the two-channel Kondo effect. In our mirror-symmetric model, there
is no further crossover and the 2CK fixed point is stable. Furthermore, by
additional calculations we have checked that 2CK is the stable fixed point
throughout the phase diagram, Fig.~\ref{skica}, for any $J_0$ and $\beta
\neq 1$, as expected.

\begin{figure}
\centering
\includegraphics[width=6cm]{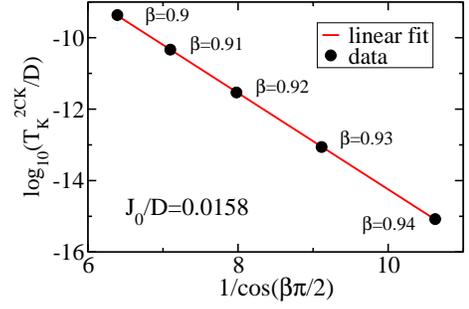}
\caption{The Kondo cross-over temperature as a function of the parameter
$\beta$ for fixed $J_0/D=0.0158$.}
\label{fig_fit}
\end{figure}

By analogy with the two-stage Kondo effect in the case of two impurities
that are coupled to a single conduction channel in a side-coupled
configuration \cite{vojta2002, cornaglia2005tsk, sidecoupled}, the
two-channel Kondo effect due to the side-coupled impurity is expected to
occur on a temperature scale that depends exponentially on the effective
exchange coupling of this impurity to the other two impurities,
$J_\mathrm{eff}$, i.e. we expect a function dependence of the form
\begin{equation}
\label{exp1}
T_K^{(2)} = T_K^{(1)} \exp\left( -\frac{1}{\rho_\mathrm{eff} J_\mathrm{eff}}
\right),
\end{equation}
where $T_K^{(1)}$ and $T_K^{(2)}$ are first-stage and second-stage Kondo
temperatures, while $\rho_\mathrm{eff}$ can be interpreted as the width of
the band of the effective local Fermi-liquid quasiparticles resulting from
the first-stage Kondo effect, which is proportional to $T_K^{(1)}$. In the
present situation, $J_\mathrm{eff} \propto \cos(\beta \pi/2)$. From the
results of calculations of the magnetic susceptibility along a line of
values of $\beta$ (from $\beta=0.9$ to $\beta=0.94$) at a fixed value of the
exchange coupling $J_0/D = 0.0158$, we extracted $T_K^\mathrm{2CK}$ using
the prescription $\mu_\mathrm{eff}(T_K^\mathrm{2CK}) = 0.07$. Linear
regression (see Fig.~\ref{fig_fit}) then gives
\begin{equation}
\label{exp2}
\log_{10} \frac{T_K^\mathrm{2CK}}{D} = -0.76 - 1.35 
\frac{1}{\cos(\beta \pi/2)}.
\end{equation}
The good agreement confirms our anticipation that Kondo screening in
multiple stages can occur whenever additional impurities are (indirectly)
weakly coupled to the continuum of electrons via other (directly coupled)
impurities. This occurs even in the case of more uncommon types of the Kondo
effect. 

\begin{figure}[htbp!]
\includegraphics[width=8cm,clip]{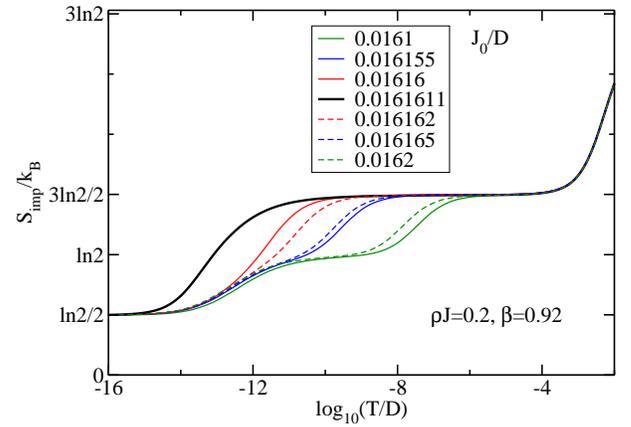}
\caption{
(Color online) Impurity cluster contribution to the entropy for a range of
values of $J_0$ at fixed $\beta$. $T_\Delta^\mathrm{2IK}$ is driven down until 
it is made equal to $T_K^\mathrm{2CK}$.
}
\label{fig2_i}
\end{figure}

\begin{figure}
\centering
\includegraphics[width=6cm]{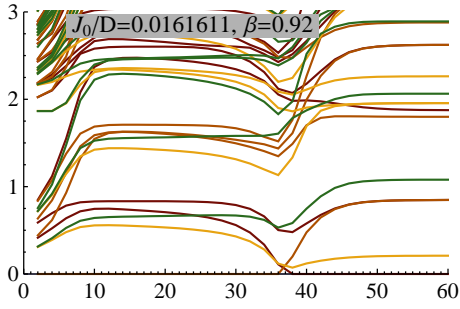}
\caption{NRG flow diagram in the case of the cross-over
from 2IK to 2CK regimes.}
\label{flow}
\end{figure}

\begin{figure}
\centering
\includegraphics[width=4cm]{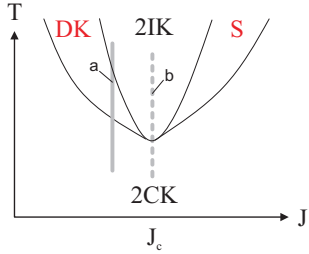}
\caption{Schematic phase diagram in the $(T,J)$ plane
for constant parameter $\beta$.}
\label{fig_phase_nfl}
\end{figure}

For $J_0/D=0.0158$, the system is already in the Fermi-liquid fixed point as
the temperature begins to decrease towards the second-stage Kondo screening.
We find, however, that the second stage Kondo screening occurs even when
after the first stage of the Kondo screening the system cannot be described
in terms of the Fermi-liquid quasiparticles, but is rather in a non-Fermi
liquid regime. In other words, the continuum of NFL excitations may also
serve as an impurity bath in a (second-stage) Kondo effect. By fine-tuning
$J_0$ around the critical value of $J^*_\mathrm{2IK}$ at fixed $\beta=0.95$,
we have managed to reduce $T_\Delta^\mathrm{(2IK)}$ below
$T_K^\mathrm{(2CK)}$, see Fig.~\ref{fig2_i}. This results in a direct
cross-over from 2IK to 2CK fixed-point on the temperature scale of $T_C$,
see the flow diagram in Fig.~\ref{flow}. It appears that the cross-over
temperature $T_C$ cannot be reduced to lower values by further tuning of the
parameter $J_0$, which implies that the phase diagram in the $(T,J)$ plane
corresponds to that sketched in Fig.~\ref{fig_phase_nfl}. The 2IK regime
does not extend down to zero temperature as would be the case at $\beta=1$
and there is a single stable 2CK fixed point at $T=0$. As the temperature is
reduced, we can either pass from 2IK to 2CK via an intermediate-temperature
Fermi-liquid phase DK or S (path $\mathrm{a}$ in Fig.~\ref{fig_phase_nfl}),
or directly from 2IK to 2CK (path $\mathrm{b}$). As $\beta$ is reduced, the
region in $(T,J)$ plane that is governed by the 2IK fixed point, becomes
smaller and eventually disappears.

\begin{figure}
\centering
\includegraphics[width=8cm]{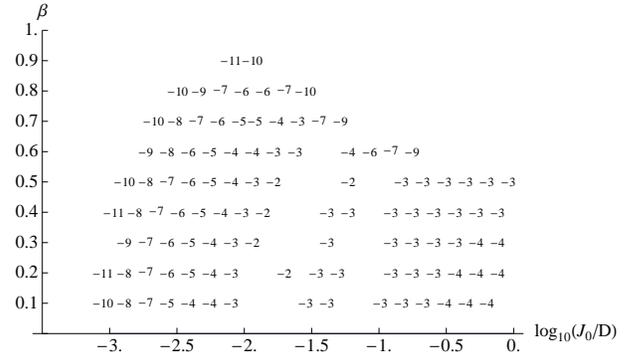}
\caption{
The cross-over (Kondo) temperatures $T_K^\mathrm{(2CK)}$ at which the system
approaches the two-channel Kondo model fixed point as a function of $J_0$
and $\beta$. The numbers shown are the integer parts of the decadic
logarithms, $|\log_{10}(T_K^\mathrm{(2CK)}/D)|$. The omitted values for
intermediate values of $J_0$ correspond to the situation where it is
difficult to determine $T_{(2CK)}$ since the cross-over occurs at relatively
high temperature. 
}
\label{t2ck}
\end{figure}

It has been shown that in the single-impurity two-channel Kondo model, the
Kondo temperature is a non-monotonic function of the Kondo exchange
interaction \cite{kolf2007}. In the weak-coupling $g=\rho_0 J_K \ll 1$
regime we have
\begin{equation}
T_K^{(wc)} \approx D e^{-1/2g+\ln(2g)+O(g)},
\end{equation}
while in the strong-coupling $g \gg 1$ regime
\begin{equation}
T_K^{(sc)} \approx D e^{-\gamma g/2-\ln(\gamma g/2)+O(1/h)}
\end{equation}
with $\gamma = 30/46$ \cite{kolf2007}. By analogy, we expect that the
temperature at which our three impurity system ultimately crosses over to
the 2CK fixed point will be a non-monotonous function of $J_0$ at any value
of $\beta$, with maximum values occurring in the parameter range where $J_0
\sim T_K^{(1)}$ with $T_K^{(1)}$ being the Kondo temperature of a single
spin-$1/2$ Kondo impurity with $\rho J_K=0.2$. This is indeed the case. In
Fig.~\ref{t2ck} we show an overview diagram of the cross-over scale as a
function of $\beta$ and $J_0$. The numbers displayed are the integer parts
of the decadic logarithm of the cross-over temperature. For very small and
very large $J_0$, $T_{2CK}$ is lower than the lowest temperature in the NRG
iteration, $\sim 10^{-12}D$. It is worth emphasizing that there is a
relatively large section of the parameter space where the cross-over occurs
at relatively high temperature. This region is a continuation to the
triangular impurity configuration of the intermediate regime found in the
linear three-impurity model discussed in Ref.~\onlinecite{flnfl3}.

We may note in conclusion to this section that the parity breaking
destabilizes the 2CK fixed point, while it is a marginal perturbation for
the 2IK fixed point. A small parity breaking of the form $J_{K,L} \neq
J_{K,R}$ or $J_{12} \neq J_{23}$ will thus lead to another crossover from
the now unstable 2CK fixed point to a stable FL fixed point (in plane P2)
which corresponds to a regular strong-couping Kondo fixed point in which the
side-coupled impurity forms a Kondo state with electrons in either left or
right channel, depending on which of the two effective exchange constants is
larger. The intermediate 2IK fixed point, on the other hand, would be
affected only little.

\section{Triangular triple quantum dot}
\label{sectqd}

A triangular triple quantum dot (TQD) model consists of three Anderson-like
impurities interconnected by electron hopping and coupled to the conduction
bands via hybridization (see also Refs.~\onlinecite{tripike} and
\onlinecite{flnfl3}). The geometry is still that displayed in
Fig.~\ref{sistem}. Various physical properties of systems with this kind of
lattice connectivity have been previously studied \cite{kuzmenko2006,
jiang2007}. In this work we focus on the Hamiltonian
\begin{equation*}
H=H_\mathrm{b}+H_\mathrm{imp}+H_\mathrm{c}
\end{equation*}
where $H_\mathrm{b}$ remains the same as in Eq.~\eqref{eq1}, while the
impurity and coupling Hamiltonians become
\begin{equation}
\begin{split}
H_\mathrm{imp} &= \sum_i (U/2) (n_i-1)^2 + 
\sum_{ij,\sigma} t_{ij} d^\dag_{i\sigma} d_{j\sigma} +
\text{H.c.}, \\
H_\mathrm{c} &= V \sum_{k\sigma} c^\dag_{L k \sigma}
d_{1\sigma} + \text{H.c.} \\
&+ V \sum_{k\sigma} c^\dag_{R k \sigma} d_{3\sigma}
+ \text{H.c.}
\end{split}
\label{eq2}
\end{equation}
Here $U$ is the on-site electron-electron repulsion, $t_{ij}$ are hopping
matrix elements, $d^\dag_{i\sigma}$ is the spin-$\sigma$ electron creation
operator on site $i$, and $n_i=\sum_\sigma d^\dag_{i\sigma} d_{i\sigma}$ is
the occupancy operator on site $i$. Finally, $V$ is the hybridization matrix
element which is assumed to be a constant independent of $k$. The
hybridization strength is then $\Gamma=\pi \rho V^2$ with the conduction
band density of states $\rho=1/(2D)$. There are no terms of the form
$\epsilon_i n_i$ which are typically included to describe the effect of gate
voltages. We instead assumed that the occupancy is near half filling, as
implied by $(U/2)(n_i-1)^2$ terms. The particle-hole symmetry is broken by
any finite hopping which makes the lattice connectivity non-bipartite and it
should be noted that the deviation from half-filling may become significant
if hopping matrix elements are large, $t_{ij} \sim U$. We are, however, more
interested in the limit of small $t_{ij} \ll U$. If the model parameters are
chosen so that $J_{ij}=4t_{ij}^2/U$ and $\rho J_K=8\Gamma/\pi U$, models
\eqref{eq1} and \eqref{eq2} have namely very similar properties as long as
$t_{ij} \ll U$. The essential difference, however, is that in model
\eqref{eq2} the electrons are allowed to tunnel from left to right
conduction leads. It is known that inter-channel charge transfer
destabilizes both the two-impurity and the two-channel Kondo-model fixed
points \cite{flnfl3, zarand2006}, therefore we expect that the ground state
of the TQD system will be Fermi-liquid for any choice of parameters. We
parameterize the tunneling matrix elements using a new quantity $\alpha$ as
\begin{equation}
\begin{split}
t_{13} &= t_0 \sin(\alpha \pi/2), \\
t_{12} = t_{23} &= t_0 \cos(\alpha \pi/2).
\end{split}
\end{equation}
Hoppings $t_{ij}$ can be related to exchange constants $J_{ij}$ only in the
limit $t_{ij} \ll U$: then the relation between $\alpha$ and $\beta$ is
given by $\sin(\beta \pi/2) \propto \sin^2(\alpha \pi/2)$.

The zero-temperature conductance through the TQD can be related to the
quasiparticle phase shifts as \cite{georges1999, pustilnik2001,
hofstetter2004, oguri2005phase, oguri2005, vzporedne2}
\begin{equation}
G=G_0 \sin^2 \left( \deltae-\deltao \right),
\end{equation}
where $G_0=2e^2/h$ is the conductance quantum. The phase shifts can be
easily extracted from the renormalization flow diagrams when the system
reaches the low-temperature stable Fermi-liquid fixed point
\cite{oliveira1981phaseshift, affleck1992, borda2003, pustilnik2001,
hofstetter2004, oguri2005, oguri2005phase, nisikawa2006, mehta2005,
zhu2006}. Phase shifts may be constrained in the presence of the
particle-hole symmetry of certain kinds \cite{affleck1995}. For $\alpha=1$,
model \eqref{eq2} is particle-hole symmetric, however the quasiparticle
scattering phase shifts are not fixed to any particular value; the stable
fixed point therefore belongs to the plane P1, see Sec.~\ref{secregimes}.
For generic $\alpha$, i.e. $\alpha \neq 1$ and $\alpha \neq 0$, the lattice
on which the Hamiltonian \eqref{eq2} is defined is not bipartite, which
immediately precludes any kind of particle-hole symmetry; the stable fixed
point must thus belong to the plane P2. Finally, the case $\alpha=0$ was
discussed in Refs.~\onlinecite{flnfl3, tripike, oguri2005}: the phase shift
in odd channel is constrained to $\pi/2$, while there is zero phase shift in
the even channel (this fixed point also belongs to the plane P2).

Properties of the double quantum dot (DQD) systems (i.e. $\alpha=1$ limit)
have been studied in a number of works using various methods
\cite{pohjola1997, aono1998, georges1999, izumida2000, boese2002, lopez2002,
aguado2003, izumida2005, karrasch2006, zarand2006, mravlje2006}. These
calculations show that the conductance of DQD goes to zero in the limit of
small $t$ (as the dots become decoupled) as well as in the limit of large
$t$ (as the electrons occupy the bonding molecular orbital); the conductance
peaks at the unitary conductance limit for $t=t^*$ such that $J_\mathrm{eff}
\sim c T_K^{(1)}$ where $J_\mathrm{eff}=4t^2/U$. The difference of phase
shifts $\Delta=\deltae-\deltao$ varies continuously and smoothly from 0 to
$\pi$ at $t$ goes from 0 to $\infty$ (it should be noted that scattering
phase shifts are defined modulo $\pi$). The fact that conductance becomes
unitary at some some value of $t$ is a simple consequence of $\Delta$ going
through $\pi/2$; what is less trivial, but not unexpected, is that this
occurs when $J_\mathrm{eff} \sim c T_K^{(1)}$.

The conductance of TQD in series, on the other hand, was found to be unitary
for all values of $t$, which is a direct consequence of the phase shift
pinning at $\deltao=\pi/2$ and $\deltae=0$ \cite{tripike, flnfl3, oguri2005,
oguri2005phase}.

At {\sl zero temperature} the variation of the phase shifts as a function of
$t_0$ and $\alpha$ might be expected to be continuous and smooth throughout
the entire parameter plane. Based on the considerations of the variation of
phase shifts as a function of $t$ in serial DQD and TQD problems, we can
make qualitative predictions about the dependence of the conductance on
$t_0$ and $\alpha$. For $t_0 \ll t^*$ we expect the conductance to drop
monotonously from $G_0$ to 0 as $\alpha$ goes from 0 to 1. For $t_0 \sim
t^*$ the conductance is $G_0$ for $\alpha=0$ and $\sim G_0$ for $\alpha=1$;
it is reasonable to assume that the conductance drops to zero for some
intermediate value of $\alpha$. Finally, for $t_0 \gg t^*$ we expect both a
conductance zero and a conductance peak for some intermediate values of
$\alpha$.

\begin{figure}
\includegraphics[width=8cm,clip]{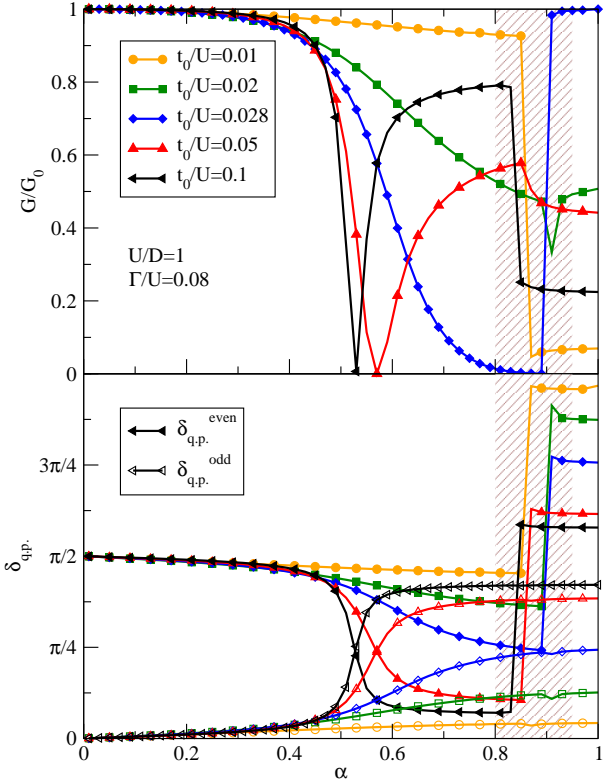}
\caption{``Zero-temperature'' linear conductance through the upper edge of
the triangular cluster and the quasiparticle scattering phase shifts as a
function of the interdot-tunneling-asymmetry factors $\alpha$ for a range of
tunneling parameters $t_0$. In the shaded box, the spectrum at the last NRG
iteration does not correspond to that of a Fermi-liquid fixed point. The
critical $t^*$ corresponding to the maximal conduction in the DQD model is
approximately $t^*=0.028 U$.
} \label{fig2_h}
\end{figure}

The results of a NRG calculation of the ``zero-temperature'' conductance are
shown in Fig.~\ref{fig2_h}. We have taken into account the spin
$\mathrm{SU}(2)$ symmetry and parity $\mathrm{Z}_2$ symmetry. In these
computations, a large value of the discretization parameter has been used,
$\Lambda=8$; up to 150 NRG iterations were performed, which corresponds to
an extremely low temperature of $T \sim 10^{-68} D$. Nevertheless, in some
parameter ranges the stable Fermi-liquid fixed point has not been reached at
this temperature. In the shaded box in Fig.~\ref{fig2_h}, the excitation
spectra at the last NRG iteration for some values of $t_0$ were not those of
Fermi liquid fixed points. Furthermore, to the right of the shaded box, in
the immediate vicinity of $\alpha=1$, spectra obtained at the last NRG
iteration correspond to Fermi-liquid fixed point in the P1 plane (see
Sec.~\ref{secregimes}), while it is known that the system should eventually
cross over to a stable Fermi-liquid fixed point in the P2 plane. These
findings can be explained by the presence of the exponentially low energy
scales when the impurity 2 is nearly decoupled; see Eqs.~\eqref{exp1} and
\eqref{exp2} for the equivalent behavior in the Kondo-like model. As
$\alpha$ goes to 1, the cross-over temperature scale $T_\Delta$ becomes
arbitrarily low, thus the ``zero-temperature'' conductance is experimentally
irrelevant.

Transport experiments are decidedly performed at some finite temperature
$T_\mathrm{exp}$. As $\alpha$ is increased, at some value $\alpha_D$ the
crossover temperature $T_\Delta$ will suddenly decrease (exponentially)
below $T_\mathrm{exp}$. For all practical purposes, the quantum dot $2$ will
then be effectively decoupled from the rest of the system and its local
moment will not be screened. It should be emphasized that $\alpha_D$ is less
than 1; at finite temperature, the local moment effectively decouples from
the rest of the system at $\alpha=\alpha_D$ even though electrons are still
able to hop on the quantum dot.

Interestingly, this behavior also implies that the conductance can be
abruptly changed by driving the crossover temperature below the experimental
temperature by relatively small variation of the gate voltages that control
$\alpha$ \cite{sidecoupled}. To our knowledge, this type of abrupt
conductance change has not yet been experimentally observed.

\begin{figure}
\centering
\includegraphics[width=8cm,clip]{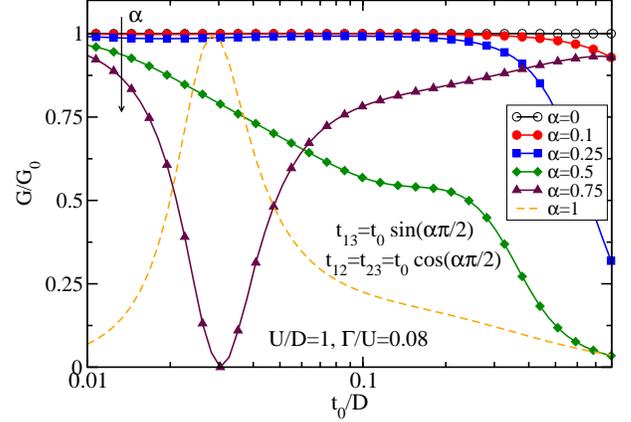}
\caption{``Zero-temperature'' linear conductance through the upper edge of
the triangular cluster as a function of the inter-dot tunneling $t_0$
for a range of tunneling-asymmetry factors $\alpha$.}
\label{fig2_g}
\end{figure}

In Fig.~\ref{fig2_g} we show the low-temperature conductance through the
triangular TQD for a range of $\alpha$ as a function of $t_0$. Particularly
interesting is how the deviation from the unitary conductance in the linear
TQD ($\alpha=0$ limit) develops when hopping between the first and the third
impurity is allowed ($\alpha \neq 0$). The differences are most significant
for large values of $t_0$; this coincides with the regime of $t_{ij} \sim U$
where the quantum dots are no longer restrained to half-filling for $\alpha
\neq 0$. The change in scattering phase shifts can thus be related to a
change in the occupancy (Friedel sum rule). For larger values of $\alpha$, 
the conductance may become zero at some value of $t_0$, see the
$\alpha=0.75$ plot in Fig.~\ref{fig2_g}. For $\alpha=1$ we recover the known
results for the conductance of the DQD system, but we must keep in mind that
for $\alpha \lesssim 1$, these results correspond to a finite temperature,
$T_\mathrm{exp} \gg T_\Delta$.

\section{Conclusion} 

The very rich phase diagram of the three-impurity two-channel system makes
this a very useful toy model to study the possible behavior of generic
two-channel quantum impurity models. We find all the fixed points familiar
from simpler impurity models and we find interesting crossovers, such as
that between the two-impurity and the two-channel Kondo model
non-Fermi-liquid fixed points.

In a related triangular triple quantum dot problem we have demonstrated that
the presence of energy scales which are extremely (exponentially) low
implies that the ``zero-temperature'' conductance is experimentally
irrelevant. In theoretical studies of impurity clusters it is thus
imperative to consider the thermal effects and to determine the temperature
scale, below which the the transport properties of the system are determined
by its ground state.

It would be interesting to extend these studies to three-impurity
three-channel models. Triple quantum dot systems with three conduction leads
can be easily manufactured today. In order to determine the transport
properties of such nanostructures in all parameter regimes (in particular in
the low temperature regime where correlation effects and Kondo physics play
a central role), an unbiased method such as NRG is required. NRG becomes
numerically highly demanding method in the case of multi-channel problems
due to the high degeneracy of quantum states that need to be considered.
While two-channel calculations are now performed routinely, only very few
three-channel calculations have been reported in the literature so far
\cite{paul1996, deleo2005, ferrero2007}. Fortunately, the transport
properties at low temperatures (if the system is near a Fermi-liquid fixed
point) can be extracted from the energy levels of the NRG eigenvalue flow
alone, which requires far less computational resources than calculations of
thermodynamic and dynamic (spectral) quantities.

\begin{acknowledgments}
The authors acknowledge support of the SRA under Grant No. P1-0044.
\end{acknowledgments}

\bibliography{paper}

\end{document}